\documentstyle[aps,prd,preprint,epsf]{revtex}
\tighten
\def\a{\alpha}
\def\b{\beta}

\begin{document}

%\preprint{.....} \draft

%\twocolumn[\hsize\textwidth\columnwidth\hsize\csname
%@twocolumnfalse\endcsname
%_________________________________________________________
%\title{Cosmic no hair: second order perturbations of
%de\,Sitter universes}
\preprint{} \draft
\title{Cosmic no-hair:\\ non-linear asymptotic stability of
de\,Sitter universe}
%__________________________________________________________
\author{Marco Bruni$^{\natural}$, Filipe C. Mena$^{\flat}$, \& Reza
Tavakol$^{\flat}$}
%___________________________________________________________
\address{
$^{\natural}$School of Computer Science and Mathematics,\\
Portsmouth University, Portsmouth PO1 2EG, U.K.\\
$^{\flat}$Astronomy Unit, School of Mathematical Sciences,\\ Queen
Mary, University of London, Mile End Road London E1 4NS, U.K.}
%________________________________________________________
%\date{July 20, 2001}
%________________________________________________________
\maketitle
%________________________________________________________
\begin{abstract}
%________________________________________________________
We study the asymptotic stability of de\,Sitter spacetime with respect 
to non-linear perturbations, by considering second order perturbations 
of a flat Robertson-Walker universe with dust and a positive 
cosmological constant.  Using the synchronous comoving gauge we find 
that, as in the case of linear perturbations, the non-linear 
perturbations also tend to constants, asymptotically in time.  
Analysing curvature and other spacetime invariants we show, however, 
that these quantities asymptotically tend to their de\,Sitter values, 
thus demonstrating that the geometry is indeed locally asymptotically 
de\,Sitter, despite the fact that matter inhomogeneities tend to 
constants in time.  Our results support the inflationary picture of 
frozen amplitude matter perturbations that are stretched outside the 
horizon, and demonstrate the validity of the cosmic no-hair conjecture 
in the nonlinear inhomogeneous settings considered here.
%____________________________________________________
\end{abstract}
\pacs{PACS numbers:  98.80.Hw,  98.80.Cq, 04.25.Nx}

%\vskip2pc]

%\newpage
%______________________________________________________
%\def\section{Introduction}
%______________________________________________________
The cosmic no-hair conjecture~\cite{Gibbons-Hawking} states -- roughly 
speaking -- that `all expanding universe models with a positive 
cosmological constant asymptotically approach the de\,Sitter 
solution'~\cite{Wald}.  Previous studies of this conjecture fall into 
two main groups: those involving special models, e.g.\ containing 
Killing vectors or special Petrov types, and those using a linear 
perturbative analysis.  Among the first group is the work of 
Wald~\cite{Wald}, who proved the conjecture for ever expanding 
homogeneous models, as well as studies of inhomogeneous 
models~\cite{Starobinsky83}.  Among the second group are studies of 
the stability of de\,Sitter with respect to linear tensor 
(gravitational wave) perturbations~\cite{Starobinsky79}, as well as 
investigations involving both linear scalar and tensor 
modes\cite{Barrow,Boucher-Gibbons}, where the scalar modes arise in 
presence of matter.  The authors of~\cite{Barrow,Boucher-Gibbons} 
studied the dynamics of linear perturbations in 
Friedmann-Lema\^{\i}tre-Robertson-Walker (FLRW) models with dust and a 
positive cosmological constant $\Lambda$, by employing a synchronous 
comoving gauge, and showed that both scalar and tensor modes tend to 
constants in time, while the vector modes were shown to 
decay~\cite{Barrow}.  This establishes the stability of de\,Sitter in 
these linear perturbative settings, but not its asymptotic stability 
~\cite{Barrow}.  It was also shown in~\cite{Boucher-Gibbons}, however, 
that there exists a coordinate transformation (valid within the 
horizon, see below) in terms of which the asymptotic metric becomes 
locally de\,Sitter in static coordinates~\cite{Hawking-Ellis}.  This 
at least partly demonstrates that one in fact has a local asymptotic 
stability, in the sense that for a freely falling observer in such a 
frame the spacetime approaches de\,Sitter in time, exponentially fast.

A number of studies have addressed this conjecture from a broader 
perspective (see e.g.~\cite{Friedrich} and Refs.\ therein), and in 
particular linear perturbations of de\,Sitter have been studied 
extensively in the context of the inflationary scenario (see 
e.g.~\cite{bi:MFB,LL} and Refs.\ therein).  The latter has been one of 
the main motivations for the interest in the cosmic no-hair 
conjecture.  Indeed inflation was devised as a mechanism to account 
for the observed large-scale homogeneity and isotropy of the universe, 
and while it is natural to assume a FLRW model to work out 
observational predictions of inflationary models such as perturbation 
spectra, it is nevertheless important to address the fundamental 
question of whether locally a FLRW geometry can arise from an open set 
of generic initial data, whose memory is lost through an inflationary 
phase.

The aims of this Letter are twofold: to extend the perturbative 
analysis of the stability of de\,Sitter to second order and to provide 
an invariant characterization of the asymptotic results of both first 
and second order perturbations.

Firstly, in order to make contact with earlier results, we study the 
asymptotic evolution of the first and second order perturbations of a 
flat FLRW model with dust plus $\Lambda$.  Asymptotically, this can be 
viewed as a de\,Sitter universe with dust perturbations.  Secondly, to 
throw more light on the asymptotic state of the perturbed universe 
model, we employ curvature and other spacetime invariants.  This 
Letter contains a short account of our results; a more detailed 
analysis will be presented elsewhere \cite{BMT2001}.

We proceed by using the formalism of \cite{MMB,MM}, extending it to
include $\Lambda$.  In particular we work in the comoving synchronous
gauge and assume vanishing vorticity.  There is no loss of generality
in the latter choice, as vorticity decays during expansion.  The line
element of the perturbed spacetime can be written as
\begin{equation}
\label{metric}
ds^2=a^2(\tau)(-d\tau^2+\gamma_{\alpha\beta}dx^\alpha
dx^\beta)\;,
\end{equation}
where $a$ is the scale factor, $\tau$ is the conformal time given by
$d\tau=dt/a(t)$ and, in usual notation
\begin{equation}
\label{perturbedm}
\gamma_{\alpha\beta}=\delta_{\alpha\beta}+\gamma^{(1)}_{\alpha\beta}+
\frac{1}{2}\gamma^{(2)}_{\a\b}\;.
\end{equation}

The full equations to be considered are: those for the evolution of 
the expansion tensor $\theta_{\alpha\beta}$
\begin{equation}
\label{evolution}
{\theta}'_{\a\b}+\theta\theta_{\a\b}+R^*_{\a\b}=
\frac{1}{2}\rho\delta_{\a\b}+\Lambda\delta_{\a\b}\;,
\end{equation}
its trace, the Raychaudhuri equation
\begin{equation}
\label{Ray}
\theta'+\theta^{\a\b}\theta_{\a\b}+\frac{1}{2}\rho=\Lambda\;,
\end{equation}
the continuity equation
\begin{equation}
\label{continuity}
\rho'+\theta\rho=0\;,
\end{equation}
and the
energy constraint equation
\begin{equation}
\label{energy}
\theta^2-\theta^{\a\b}\theta_{\a\b}+R^*=2(\rho+\Lambda)\;,
\end{equation}
where $\rho$ is the matter-density, 
$\theta^\a_\b=u^\a_{;\b}=\frac{1}{2}\gamma^{\a\delta}{\gamma}'_{\delta\b}$, 
$R^*$ is the 3-Ricci scalar and the prime denotes the derivative with 
respect to $\tau$.  Since we are interested in the asymptotic behavior 
of the perturbations, we shall assume that the background evolution 
has already reached the asymptotic regime, with the $\Lambda$ term 
dominating in the Friedmann equation.  The Friedmann equation can in 
this case be solved to give
\begin{equation}
\label{scale}
a(\tau)=-\frac{H^{{-1}}}{\tau}\;, ~~~~ H=\sqrt{\frac{\Lambda}{3}}\;,
\end{equation}
where the conformal time $\tau$ is negative and
tends to zero as $a$ goes to infinity.

We start by considering the first order perturbations since they act 
as source terms in the second order perturbation equations.  In the 
linear theory the scalar and tensor modes are independent and for an 
irrotational spacetime one can set the vector modes equal to zero.  
The evolution equation for the scalar perturbations is obtained by 
combining (\ref{evolution}), (\ref{continuity}) and (\ref{energy}) and 
can be written in terms of the density contrast 
$\delta=(\rho-\rho_b)/{\bar \rho}$ in the form
\begin{equation}
\label{delta}
\delta''+\frac{a'}{a}\delta' -\frac{1}{2} a^2 \bar{\rho} \delta =0\;,
\end{equation}
where ${\bar \rho}=\rho_0/a^3$ is the background density and $\rho_0$ 
is a constant.  Note that the last term of this equation is 
asymptotically negligible and therefore the asymptotic solution of 
(\ref{delta}) has two modes, one constant and the other decaying.  
Expanding the full solution of this equation around $\tau=0$ one 
obtains up to second order in $\tau$
\begin{equation}
\label{delta-asymp}
\delta(\tau,{\bf x})\approx C_1({\bf
x})\left(\frac{\Gamma(\frac{2}{3})}{A^{\frac{2}{3}}\pi}+
\frac{1}{6}\frac{3^{\frac{1}{6}}A^{\frac{2}{3}}}{\Gamma(\frac{2}{3})}\tau^2
\right)
+C_2({\bf x})\tau^2\;,
\end{equation}
where $A^2=\frac{1}{2}\rho_0 H$; $C_1$ and $C_2$ are arbitrary spatial 
functions related to the initial data.  The scalar metric 
perturbations are related to $\delta$~\cite{MMB} and also tend to 
constants in time~\cite{BMT2001}.

The tensor perturbation $\pi_{\a\b}$ is the transverse traceless part 
of $\gamma^{(1)}_{\a\b}$ and is gauge-invariant.  The asymptotic form 
of the equation for $\pi_{\a\b}$ is similarly obtained from the 
linearization of the evolution equations, and for each Fourier mode 
$k$ is given by
\begin{equation}
\pi^{\prime\prime}_{\a\b}-\frac{2}{\tau}\pi^{\prime}_{\a\b}+k^2\pi_{\a\b}=0
\;.
\end{equation}
This equation can be solved to give a solution for each mode $k$
\begin{equation}
\label{durban}
\pi_{\a\b}=a_{\a\b}(\sin{k\tau}-k\tau\cos{k\tau})+b_{\a\b}(\cos{
k\tau}+k\tau\sin{k\tau})\;,
\end{equation}
where $a_{\a\b}=a_{\a\b}({k})$ and $b_{\a\b}=b_{\a\b}({k})$.
A formal solution $\pi_{\a\b}({\bf x},\tau)$
can then be obtained in terms of a Fourier integral,
which can asymptotically be approximated by
\begin{equation}
\label{pi-asymp}
\pi_{\a\b}\approx\int^\infty_0
\left( b_{\a\b}\left(1+\frac{k^2\tau^2}{2}\right) +
\frac{a_{\a\b}k^3\tau^3}{3}\right)e^{-i{\bf kx}} d{\bf
k}\;.
\end{equation}
Thus both the first order scalar and tensor
perturbations (\ref{delta-asymp}) and (\ref{durban})
asymptotically approach constants in time.

The analysis so far confirms known results 
\cite{Starobinsky79,Barrow,Boucher-Gibbons} and sets the stage for 
proceeding to the second order.  Before doing so, it is however 
worthwhile taking a brief look at some first order gauge-invariant 
quantities.  The Electric Weyl tensor satisfies~\cite{BDE}
\begin{equation}
\label{ewayl}
a ^{(3)}\nabla_{b} E^{b}_{a} =\case{1}/{3}\bar{\rho} {\cal D}_{a} ~~~
\Leftrightarrow ~~~ 2\case{k^{2}}/{a^{2}}
\Phi_{H}=\bar{\rho}\varepsilon_{m}\;,
\end{equation}
where the equivalence with the equation on the right hand side (Eq.\ 
(4.3) in Bardeen~\cite{bi:bardeen}) is in Fourier space; ${\cal 
D}_{a}$ is the comoving fractional density gradient of density, a 
covariant gauge-invariant measure of inhomogeneity~\cite{Ellis-Bruni}, 
and $\varepsilon_{m}$ is the equivalent Bardeen variable~\cite{BDE}.  
Now both these quantities evolve exactly as the density perturbation 
in the comoving gauge above, and thus tend to constant values in time 
in the present context.  The simple Eq.\ (\ref{ewayl}) is the 
relativistic perturbation equivalent of the Poisson equation, and 
Bardeen's potential $\Phi_{H}$ is the gauge-invariant analogue of the 
Newton's potential~\cite{BDE}.  This equation carries a great deal of 
useful information.  Firstly it shows that in the absence of matter, 
the Electric Weyl tensor is transverse, so that there are no scalar 
modes in this case, $\Phi_{H}=0$, but only gravitational waves (cf.\ 
\cite{higuchi}.  Secondly it shows that $\Phi_{H}$ decays as $\sim 
a^{-1}$ when the density perturbation is frozen, with a corresponding 
decay in (the scalar part of) $E^{b}_{a}$.  Finally $\Phi_{H}$ 
directly corresponds to the metric perturbation variable in the 
longitudinal gauge~\cite{bi:MFB}, which therefore also decays (cf.\ 
\cite{MMB}).

Concerning the second order perturbations, the spatial metric in 
(\ref{metric}) can be written as
\begin{equation}
\label{perturbed}
\gamma^{(2)}_{\a\b}=-2\phi^{(2)}\delta_{\a\b}+\chi^{(2)}_{\a\b\;},
\end{equation}
where $\phi^{(2)}$ and $\chi^{(2)}_{\a\b}$ denote respectively the 
trace and trace-free parts.  The usual procedure is then to substitute 
(\ref{perturbed}) and (\ref{scale}) in the field equations, using a 
similar procedure as in the linear case.  Keeping the lowest order 
terms of (\ref{delta-asymp}) and (\ref{pi-asymp}) on the right hand 
side of the Raychaudhuri equation we obtain
\begin{equation}
\label{2ndphi}
{\phi_{\scriptscriptstyle}^{(2)}}''-
{1 \over \tau}
{\phi_{\scriptscriptstyle}^{(2)}}'+\frac{1}{2}\sqrt{\frac{\Lambda}{3}}
\rho_0\tau\phi^{(2)}=F({\bf x})\tau+O(\tau^2)\;,
\end{equation}
where the source term $F$ is a spatial function (given in 
\cite{BMT2001}) that depends quadratically on first order 
perturbations, and in particular from the four initial data quantities 
$C_1, C_2, a_{\a\b}$ and $b_{\a\b}$.  The solutions to (\ref{2ndphi}) 
have the form
\begin{equation}
\phi^{(2)}=\frac{F({\bf x})}{A^2}+
\tau
C_3({\bf x})J_{\frac{2}{3}}\left(\case{2A\tau^{3/2}}/{3}\right)+
\tau
C_4({\bf x})Y_{\frac{2}{3}}\left(\case{2A\tau^{3/2}}/{3}\right),
\end{equation}
with $C_3$ and $C_4$ are arbitrary spatial functions.  The asymptotic 
behavior of this solution can be obtained by expanding around $\tau=0$ 
to give, up to second order in $\tau$
\begin{equation}
\label{phi2-expansion}
\phi^{(2)}\approx \frac{F({\bf x})}{A^2}+C_3({\bf
x})\left(\frac{\Gamma(\frac{2}{3})}{A^{\case{2}/{3}}\pi}\right)
+
C_4({\bf x}) \tau^2\;.
\end{equation}
Now, assuming standard second order initial conditions~\cite{MMB}
$
\phi^{(2)}(\tau_0,{\bf x})=0$ and $\phi^{(2)'}(\tau_0, {\bf x})=0
$
we find
\begin{equation}
C_3({\bf x})=\frac{-F/A^2}{(\frac{B_1}{B_2}-B_2)\tau^2_0-B_1}\;,
~~C_4({\bf x})=\frac{-F/A^2}{1+(\frac{B_2}{B_1}-1)\tau^2_0}\;,
\end{equation}
where $B_1=\Gamma(\frac{2}{3})/(A^{\frac{2}{3}}\pi)$ and 
$B_2=3^{\frac{1}{6}}A^{\frac{2}{3}}/(6\Gamma(\frac{2}{3}))$.  
Therefore, $\phi^{(2)}$ asymptotically approaches a constant value in 
time which depends on $\tau_0$, $C_1$, $C_2$, $a_{\a\b}$ and 
$b_{\a\b}$.  Furthermore, from the momentum and energy constraints, we 
obtain up to second order in $\tau$
\begin{equation}
\label{w}
\chi^{(2)\a}_{\b,\a}(\tau,{\bf x})\approx Q_\b ({\bf x})\;,
~~
\chi^{(2)\a\b}_{,\a\b}(\tau,{\bf x})\approx G({\bf x})\;,
\end{equation}
where $Q_\b$ and $G$ are spatial functions given in
\cite{BMT2001}. Finally,  substituting (\ref{w}) in the asymptotic
evolution equation for the second order trace-free perturbations
we obtain
\begin{equation}
\chi^{(2)''}_{\a\b}-\frac{2}{\tau} \chi^{(2)'}_{\a\b}+k^2\chi^{(2)}_{\a\b}=
 S_{\a\b}(k)+O(\tau)\;,
\end{equation}
where $S_{\a\b}$ is a function of $k$ given in \cite{BMT2001}, and, as 
$F$ above, is quadratic in first order quantities.  This equation can 
be solved to give
\begin{eqnarray}
\chi^{(2)}_{\a\b}=
\frac{S_{\a\b}(k)}{k}&+&c_{\a\b}(k\tau\cos{k\tau}-\sin{k\tau})+\\
&&d_{\a\b}(k\tau\sin{k\tau}+\cos{k\tau})\;,\nonumber
\end{eqnarray}
where $c_{\a\b}$ and $ d_{\a\b}$ are functions of $k$, which can be
expanded around $\tau=0$ and formally integrated to give
\begin{equation}
\chi^{(2)}_{\a\b}(\tau,{\bf x})\approx
\int^\infty_0\left(\frac{S_{\a\b}(k)}{k}+
d_{\a\b}\left (1-
\frac{k^2\tau^2}{2} \right ) \right) e^{-i{\bf k}{\bf x}}d{\bf
k}\;.
\end{equation}
As a result, as $\tau\to 0$, $\chi^{(2)}_{\a\b}$ tends to a constant
in time which depends on the asymptotic values of $\phi^{(2)}$ and
$d_{\a\b}$.  These values are related to the first
order initial free data by setting the initial conditions $
\chi^{(2)}_{\a\b}(\tau_0,k)=0$ and $\chi^{(2)'}_{\a\b}(\tau_0,k)=0 $,
which result in
\begin{equation}
c_{\a\b}(k)=\frac{-S_{\a\b}\tau_0}{\frac{k^4\tau_0^4}{6}-k^3\tau_0^2}\;,~
~
d_{\a\b}(k)=\frac{-S_{\a\b}}{1+\frac{k\tau_0^2}{6}}\;.
\end{equation}
So at second order there are still four independent quantities, namely 
$C_1, C_2, a_{\a\b}$ and $b_{\a\b}$, which correspond to the first 
order free initial data.  Second order density perturbations can be 
computed once the metric is given~\cite{BMT2001}.

In this way, we have shown that the second order scalar and tensor
perturbations asymptotically approach constant values in time that are
dependent on the initial conditions.  This demonstrates that the
asymptotic behavior of the first order perturbations is stable with
respect to the second order nonlinear perturbations.

Now as was the case for the first order perturbations (see
\cite{Barrow,Boucher-Gibbons}), these asymptotic constant values
cannot be removed globally by a gauge transformation. However,
this can be done locally: by rescaling the time coordinate, one
can write the asymptotic form of (\ref{metric}) for the case of
the second order perturbations as
\begin{equation}
\label{asymp}
ds^2=-dt^2+e^{2\sqrt{\frac{\Lambda}{3}}t}A_{\alpha\beta}(x,y,z)dx^\alpha
dx^\beta,
\end{equation}
where $A_{\a\b}$ is the asymptotic form of the spatial metric 
$\gamma_{\a\b}\approx A_{\a\b}+B_{\a\b}\tau^2$, and both $A_{\a\b}$ 
and $B_{\a\b}$ are spatial quantities corresponding to the sum of the 
first and second order asymptotic perturbation contributions.  Then as 
shown in \cite{Boucher-Gibbons}, there exists a coordinate change 
(valid inside the event horizon of a freely falling observer at 
$y^\alpha=0$) given by
\begin{eqnarray}
&&y^\alpha=e^{Ht}x^\alpha\nonumber\\ &&e^{HT}=\frac{e^{Ht}}
{\sqrt{1-H^2y^2}},
\end{eqnarray}
where $y^2=y_1^2+y_2^2+y_3^2$, which allows the line
element (\ref{asymp}) to be asymptotically written as
\begin{equation}
ds^2=-(1-H^2r^2)dT^2+(1-H^2r^2)^{-1}dr^2+r^2d\Omega^2,
\end{equation}
which represents the de\,Sitter solution in static coordinates 
\cite{Hawking-Ellis}.  In this sense, as in the case of first order 
perturbations, a freely falling observer will locally see the space 
asymptotically approach de\,Sitter in time.

This result is somehow unsatisfactory, as it relies on a specific 
frame and does not provide a direct insight into the fate of physical, 
observable gauge-invariant quantities~\cite{BS} (see also 
\cite{MMB,BMMS}).  Therefore, in order to shed more light on this 
behavior, we make use of curvature~\cite{carminati} and other 
invariants in order to characterize the asymptotic nature of the 
spacetime.  The conformal electric and magnetic parts of the Weyl 
tensor can be written as
\begin{eqnarray}
\tilde
E^\a_\b=a^2E^\a_\b&=&\tilde\theta\tilde\theta^\a_\b+\frac{a'}{a}
\tilde\theta^\a_\b+
-\tilde\theta^\a_\gamma
\tilde\theta^\gamma_\b+\\
&&\frac{1}{3}\delta^\a_\b(\tilde\theta^\mu_\nu\tilde\theta^\nu_\mu-
\tilde\theta^2-\frac{a'}{a}\tilde\theta)
+\tilde R^\a_\b-\frac{1}{3}\delta^\a_\b \tilde R\;,\nonumber\\
\tilde H^\a_\b=a^2H^\a_\b&=&\frac{1}{2}\gamma_{\b\mu}(\eta^{\mu\gamma\delta}
\sigma^\a_{\gamma;\delta}+\eta^{\a\gamma\delta}\sigma^\mu_{\gamma;\delta})\;
,
\end{eqnarray}
where $\tilde\theta^\a_\b=a\theta^\a_\b-\frac{a'}{a}\delta^\a_\b$,
$\tilde R^\a_\b=a^2R^{*\a}_\b$,
$\sigma^\a_\b=\tilde \theta^\a_\b-\frac{1}{3}\delta^\a_\b\tilde\theta$
and $\eta^{\a\b\gamma}=\epsilon^{\a\b\gamma}/\sqrt{\gamma}$.
It is then possible to show that
asymptotically $\tilde\theta^\a_\b\to 0$,
$\tilde R^\a_\b\to const$ (cf. \cite{MMB} for the
expression
for
$\tilde R^{*\a}_\b$ perturbed to second order),
$\tilde H^\a_\b\to 0$ and
$\tilde E^\a_\b\to const$, and
therefore we have $\theta^\a_\b\to \sqrt{\Lambda/3}$, $\sigma^\a_\b\to 0$,
$R^{*\a}_\b\to 0$,
$E^\a_\b \to 0$ and $H^\a_\b \to 0$. These in turn give
\begin{eqnarray}
\label{inv1}
&& \theta^2\to \Lambda\;, ~~~ H^2\to 0\;,~~R\to4\Lambda\;, \\
\label{inv2}
&& \sigma^2 \to 0\;,~~~E^2\to 0\;,~~~R^*\to 0\;.
\end{eqnarray}
Now given that in the case of dust acceleration is zero and vorticity 
vanishes asymptotically, then Eqs.\ (\ref{inv1})-(\ref{inv2}) are 
sufficient to prove invariantly that, locally, the nonlinearly 
perturbed spacetime approaches de\,Sitter.

To conclude, we have studied the asymptotic evolution of a dust plus 
$\Lambda$ spacetime with vanishing vorticity, by considering second 
order perturbations of a dust plus $\Lambda$ FLRW universe which is 
asymptotically de\,Sitter.  We have proved the local asymptotic 
stability of the latter, using curvature and other spacetime 
invariants.  This is a general second order gauge-invariant result for 
the case of irrotational dust with a positive cosmological constant, 
as guaranteed by the vanishing of $\sigma^\a_\b$, $R^{*\a}_\b$, 
$E^\a_\b$ and $H^\a_\b$ in the background {\it and}\ (asymptotically) 
at first order~\cite{MMB,BS,BMMS}.  It is well known that vorticity 
decays in an expanding perfect fluid with equation of state $p=w\rho$, 
$w<2/3$, and one can prove that first order scalar perturbations decay 
for $-1<w<2/3$ in an asymptotically de\,Sitter 
background~\cite{BMT2001}.  Gravitational wave perturbations are 
affected by the equation of state only through the 
background~\cite{bi:bardeen}, and therefore their asymptotic evolution 
and their contribution to invariants is as described here.  Therefore 
our results can in this sense be said to be more general.

We have demonstrated the validity of the cosmic no-hair conjecture in 
the nonlinear inhomogeneous settings considered here.  This, together 
with previous exact and perturbative results, also supports the 
picture that, starting from inhomogeneous initial conditions, the 
universe as a whole may consist of homogeneous isotropic patches that 
have emerged from inflationary phases, and others that have undergone 
recollapse.

Non-linearities in the relativistic cosmology of the early and late 
universe have recently been investigated in a number of settings (see 
e.g.\ \cite{nonlin,MMB} and Refs.\ therein).  We believe that such 
non-linear approximate methods are of great potential value in 
tackling many interesting problems in this field.

\vspace{.1in} FCM thanks M. MacCallum for interesting comments, CMAT, 
U.Minho for support, FCT (Portugal) for grant PRAXIS XXI BD/16012/98, 
and the Relativity and Cosmology group at Portsmouth for warm 
hospitality.  MB thanks Bruce Bassett for a useful remark and MB and 
RT would like to thank the `Mathematical cosmology programme', Erwin 
Schr\"odinger Institute (Vienna) for hospitality while this work was 
finalised.

%___________________________________________________________________________
\end{document}